\documentclass[aps,prb,floats,twocolumn]{revtex4}
 \usepackage{epsfig}
\usepackage{graphics}
\usepackage{amsmath}
\usepackage{amssymb}
\begin{document}
\title{Mechanism of temperature dependence of the magnetic anisotropy energy in ultrathin Cobalt and Nickel films}
\author{J. Kienert$^{1}$, S. Schwieger$^{2,*}$, K. Lenz$^{3}$, J. Lindner$^{4}$, K. Baberschke$^{3}$, and W.~Nolting$^{1}$}
\affiliation{
$^{1}$Lehrstuhl Festk{\"o}rpertheorie, Institut f{\"u}r Physik, Humboldt-Universit{\"a}t zu Berlin, 
  Newtonstr. 15, 12489 Berlin, Germany\\
$^{2}$Technische Universit\"at Ilmenau, Theoretische Physik I,
           Postfach 10 05 65, 98684 Ilmenau, Germany\\
$^{3}$Institut f\"ur Experimentalphysik, Freie Universit\"at 
Berlin, Arnimallee 14, 14195 Berlin, Germany\\
$^{4}$Fachbereich Physik, Experimentalphysik-AG Farle, 
Universit\"at Duisburg-Essen, Lotharstr. 1, 47048 Duisburg, Germany\\
$^{*}$Corresponding author, Stephan.Schwieger@TU-Ilmenau.de}
\begin{abstract}
Temperature dependent FMR-measurements of Ni and Co films are analysed
using a microscopic theory for ultrathin metallic systems. The mechanism
governing the temperature dependence of the magnetic anisotropy 
energy is identified and discussed. It is reduced with increasing
temperature. This behavior is found to be solely caused by magnon excitations.
\end{abstract}
\maketitle
\section{Introduction}\label{}
Research on ultrathin films has been growing considerably over the last two decades due to their technical importance and the 
increasing ability to grow high-quality film samples. When dealing with systems of reduced dimensionality it is important to take into account the 
influence of magnetic anisotropies. Formally being a prerequisite for finite temperature magnetism \cite{MW66} they also play a crucial role in reorientation 
transitions as functions of film thickness, temperature, or an external magnetic field \cite{SKN05a}.

In this work we present a microscopic quantum mechanical model to describe the temperature dependence of the magnetic anisotropies of ultrathin 
Ni and Co films. As the dominant anisotropy terms in these transition metals we take into account a second order lattice anisotropy and dipolar coupling. 
We are able to fit experimental data obtained by the ferromagnetic resonance technique (FMR) \cite{LB03,L03} with our model, which is based on a Heisenberg 
exchange between localized magnetic moments. The main result will be that the $T$-dependence of the anisotropy is solely due to magnon excitations rather 
than to other mechanisms as, e.g., thermal expansion or phononic interactions. The isolation of the magnon effect is of course not possible 
when describing the films with the classical (T=0-) Landau-Lifshitz equations in which the (effective) anisotropy parameters have to be fitted at each given 
temperature \cite{L03}. 
Temperature is taken into account explicitly by the so-called stochastic Landau-Lifshitz-Gilbert (LLG) equation \cite{Gar97}. However we will introduce in 
the following an alternative approach based on a quantum mechanical description of the film system and an explicite consideration of 
magnon excitations using a Heisenberg model containing atomistic anisotropy terms.

\section{Theoretical description}\label{}

The Hamiltonian of our microscopic model reads
\begin{eqnarray}
H = &-&\sum_{ij}J_{ij}\mathbf{S}_{i}\mathbf{S}_{j} \label{hamilt}
 - \sum_{i}g_{J}\mu_{B}\mathbf{B}_0\mathbf{S}_{i}-\\
 &-&\sum_{i}K_{2}S_{iz'}^2 + \sum_{ij}g_{0}\left(\frac{1}{r_{ij}^3}\mathbf{S}_{i}\mathbf{S}_{j} - \frac{3}{r_{ij}^5}(\mathbf{S}_{i}\mathbf{r}_{ij})(\mathbf{S}_{j}\mathbf{r}_{ij})\right)\nonumber
\end{eqnarray}
The first term describes Heisenberg coupling $J_{ij}$ between magnetic spin moments $\mathbf{S}_{i}$ at the sites of a monolayer. It represents the largest 
energy scale in the problem and is responsible for the magnetism in the system. A film thickness beyond monolayer is effectively absorbed into the nearest 
neighbor exchange parameter $J$ to which we restrict ourselves. $J$ is chosen such that the monolayer magnetic moment equals that of the multilayer film at room temperature ($T_{C}^{Ni_{7}}=410$ K, $T_{C}^{Co_{2}}=400$ K). 
The second term contains an external magnetic field $\mathbf{B}_0$ in arbitrary direction with the Land{$\rm\acute e$} factor $g_{J}$ and the Bohr magneton $\mu_{B}$. 
The third and fourth term constitute lattice anisotropy and dipolar interaction, respectively, the latter leading to shape anisotropy. $K_{2}$ and $g_{0}$ are 
microscopic anisotropy parameters, $S_{iz'}$ is 
the $z'$-component of $\mathbf{S}_{i}$ perpendicular to the film plane, and $\mathbf{r}_{ij}$ is the vector between lattice sites $i$ and $j$. The shape anisotropy 
favors in-plane orientation and the lattice anisotropy can favor in-plane ($K_{2}<0$) or out-of-plane ($K_{2}>0$) orientation of the magnetization.

The main idea of the method that we used in order to solve (\ref{hamilt}) is discussed in Ref. \cite{SKN05a}. The exchange terms are decoupled using 
the standard Tyablikov (RPA) approximation. The crucial point is to find a reasonable decoupling of the lattice anisotropy terms in the equation of 
motion for the spin Green function $\langle \langle S_{i}^{+};S_{j}^{-}\rangle \rangle$ given an {\em arbitrarily} oriented external magnetic field. 
This problem is solved by performing a coordinate transformation $(x',y',z')\rightarrow(x,y,z)$. More precisely one self-consistently rotates the initial 
coordinate system defined by a $z'$-axis parallel to the film normal such that the $z$-axis of the new 
reference frame has the direction of the magnetization. Then an Anderson-Callen decoupling to the $K_{2}$-terms is applied \cite{AC64}. The results for the 
magnetization components obtained by this approach compare very well with corresponding QMC calculations \cite{SKN05a}. 
We have improved the theoretical treatment in the meantime by taking into account additional Green functions, namely all combinations of 
$\langle \langle S_{i}^{+,-};S_{j}^{+,-}\rangle \rangle$, as was proposed in Ref. \cite{PPS05}. This improved theory ensures correct 
softening properties of the uniform spin wave mode and also agrees nicely with the QMC results.
Furthermore the dipole term in (\ref{hamilt}) is treated in 
the RPA approximation and only the uniform (${\bf q}\rightarrow 0$) contribution is considered as the non-uniform terms are negligible compared to contributions 
from the much larger Heisenberg exchange.
\begin{figure}[t]
\includegraphics[width=0.9\linewidth]{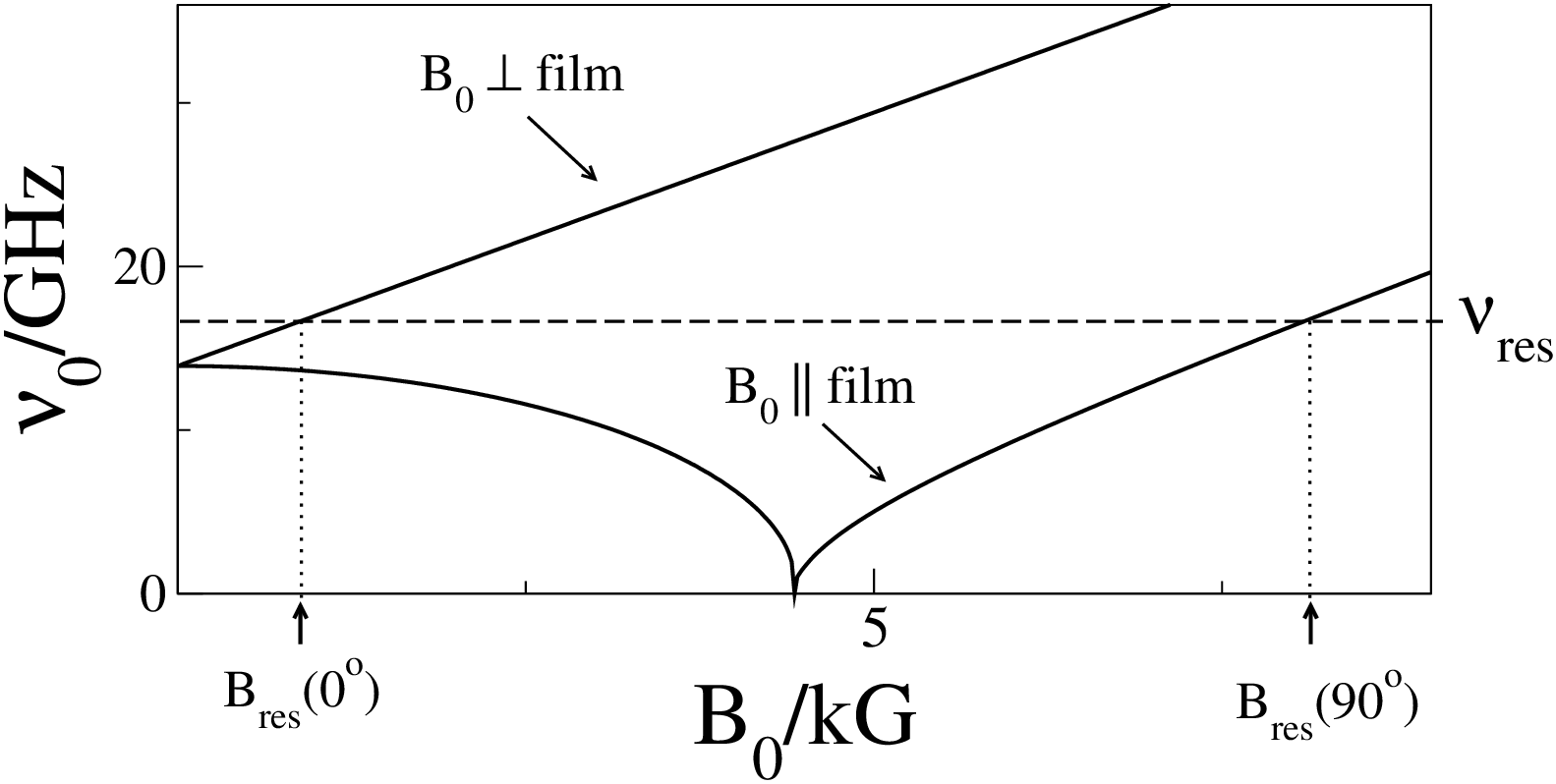}
\caption{Resonance frequency as a function of the external magnetic field applied parallel ($\theta_{B_{0}}=90^{\circ}$) and perpendicular 
($\theta_{B_{0}}=0^{\circ}$) to the film plane (monolayer). The resonance fields can be detected by an FMR experiment. Parameters: $T=0,~K_{2}=10\mu_{B}$kG, 
$g_{0}=0$.}
\label{fig1}
\end{figure}
\begin{figure}[t]
\includegraphics[width=0.9\linewidth]{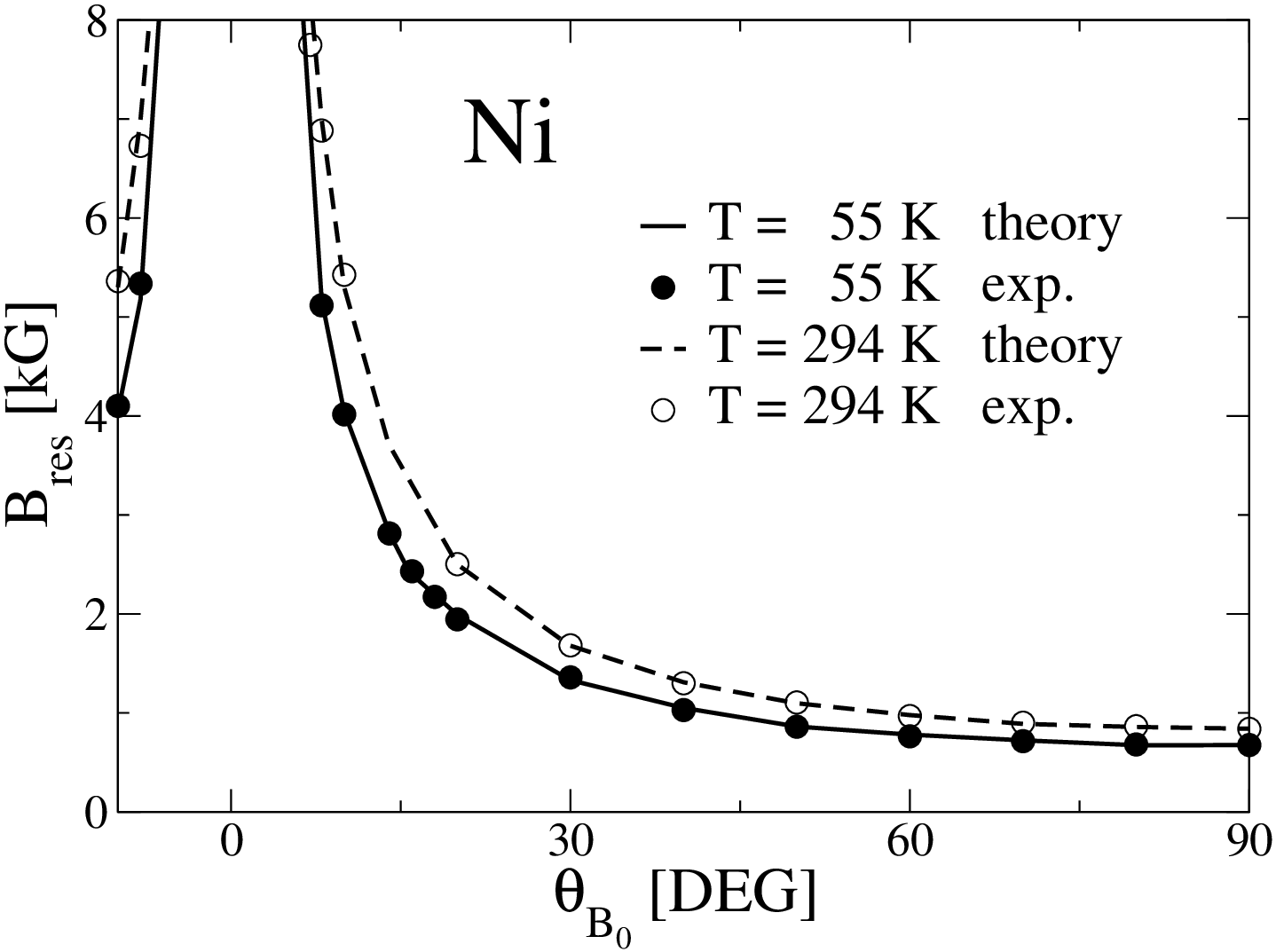}
\caption{Resonance field of Nickel at $T=55$K and at $T=294$K as a function of the orientation of the external magnetic field. Circles: experimental 
data from Ni$_7$/Cu(001). Lines connect theoretical fit points. $S$=1, $K_2=3.0\mu_B\rm kG$, $g_0=4.5\mu_B\rm kG$, $J=30$meV.}
\label{fig2}
\end{figure}

A more detailed and general account of our method (e.g. the explicit extension to the multilayer case) will be presented elsewhere. We summarize only the 
essential output here: solving for the spin Green functions yields 
weights $\chi_{\alpha}({\bf q})$ and excitation energies $E_{\alpha}({\bf q})=\hbar\omega_{\alpha}({\bf q})$ which in turn give the average magnon occupation 
number
\begin{eqnarray}
        \varphi(T) = \frac{1}{N}\sum_q \sum_\alpha \frac{\chi_{\alpha}(q)}{e^{\beta E_\alpha(q)}-1}\;.
\end{eqnarray}
The two terms of the sum over $\alpha$ describe the single-magnon excitations of the system for a given wave vector $q$, namely magnon creation and magnon 
annihilation. The magnetization (in the rotated frame) can then be computed from
\begin{equation}
\langle S_{z}\rangle = \frac{(1+\varphi)^{2S+1}(S-\varphi)+\varphi^{2S+1}(S+1+\varphi)}{(1+\varphi)^{2S+1}-\varphi^{2S+1}}\;.
\end{equation}
Our theory therefore allows for a self-consistent determination of the magnetization, i.e. temperature-dependent calculations. In addition the self-consistent determination of the rotation angle is achieved by requiring that $S_{z}$ be a constant of the motion, 
$\frac{dS_{z}}{dt}=0$, within the decoupling approximations we use in our theory. We point out that by properly rescaling the parameters our equations 
at $T=0$ can actually be shown to reduce to the Landau-Lifshitz equations. A discussion of the important differences that appear for finite temperatures will be 
given elsewhere.  

In our theory the anisotropies $K_{2}$ and $g_{0}$ influence the system solely via the {\em effective anisotropy} given by the 
temperature-dependent term 
\begin{eqnarray}
\label{Keff}
{\tilde K}_{2}(T)= \langle S_{z}\rangle(T) \left(2K_{2}C(T)-Dg_{0}\right)\;,\\
C(T)=1-\frac{(S(S+1)-\langle S_{z}^2\rangle(T))}{2S^2}\;.
\end{eqnarray}
Here the $T$-independent quantity $D$ is some number depending on the lattice geometry. Note that our effective anisotropy ${\tilde K}_{2}$ is identical to the 
quantity $M_{eff}$ commonly used within a Landau-Lifshitz description of FMR experiments \cite{LB03}. Furthermore we exploit 
$\langle S_{z}^2\rangle(T) = S(S+1)- \langle S_{z} \rangle(T) (1+2\varphi(T))$. The temperature dependence of 
${\tilde K}_{2}$ thus goes beyond a mere proportionality to $\langle S_{z}\rangle(T)$ due to the occurence of the higher order $T$-dependent correlation function 
$\langle S_{z}^2\rangle(T)$.
\begin{figure}[t]
\includegraphics[width=0.9\linewidth]{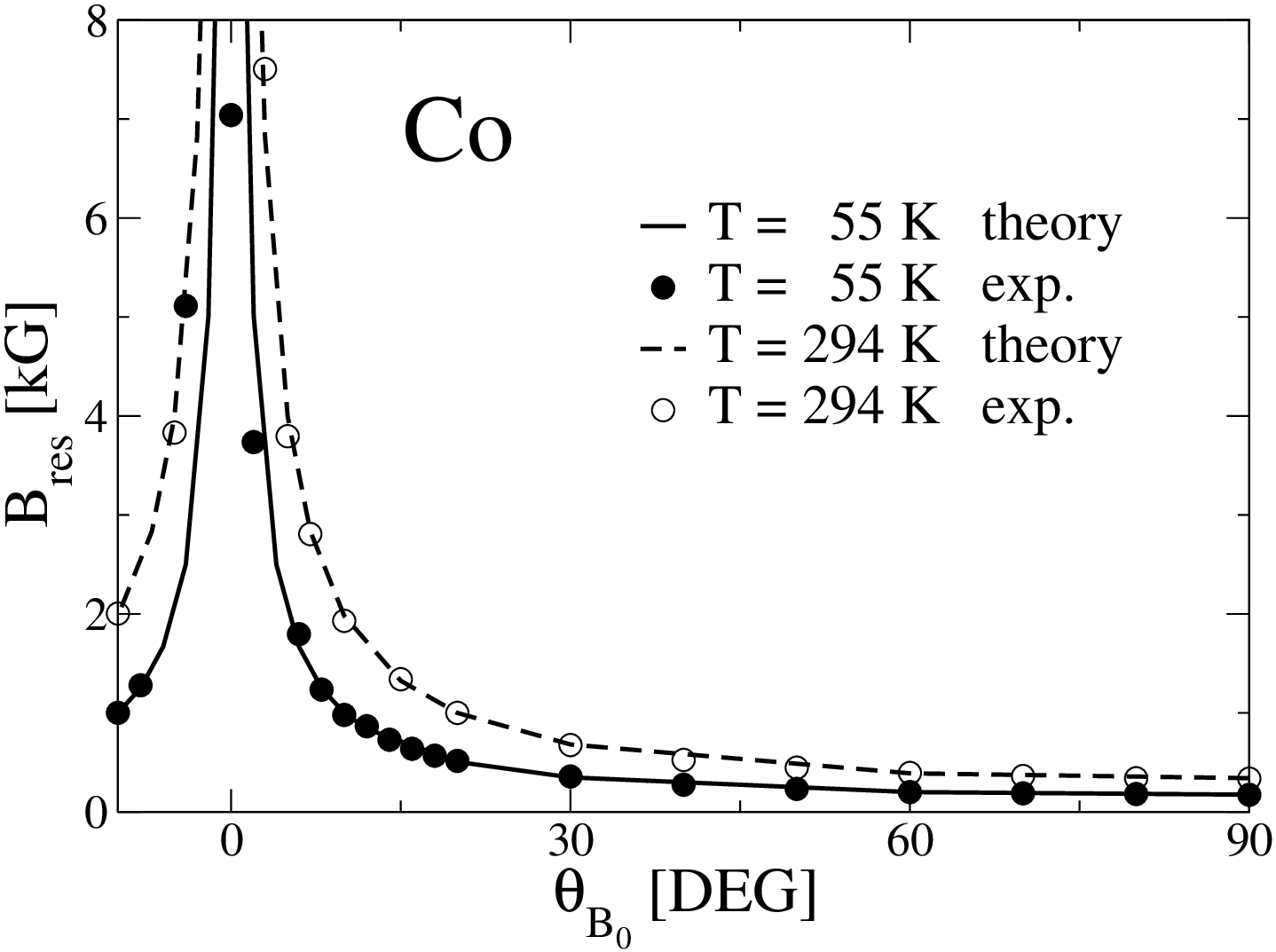}
\caption{Same as in Fig. 2 for Cobalt (Cu$_9$/Co$_2$/Cu(001)). Parameters: $S$=2.5, $K_2=-20.25\mu_B\rm kG$, $g_0=1.95\mu_B\rm kG$, $J=4.1$meV.}
\label{fig3}
\end{figure}
\begin{figure}[t]
\includegraphics[width=0.85\linewidth]{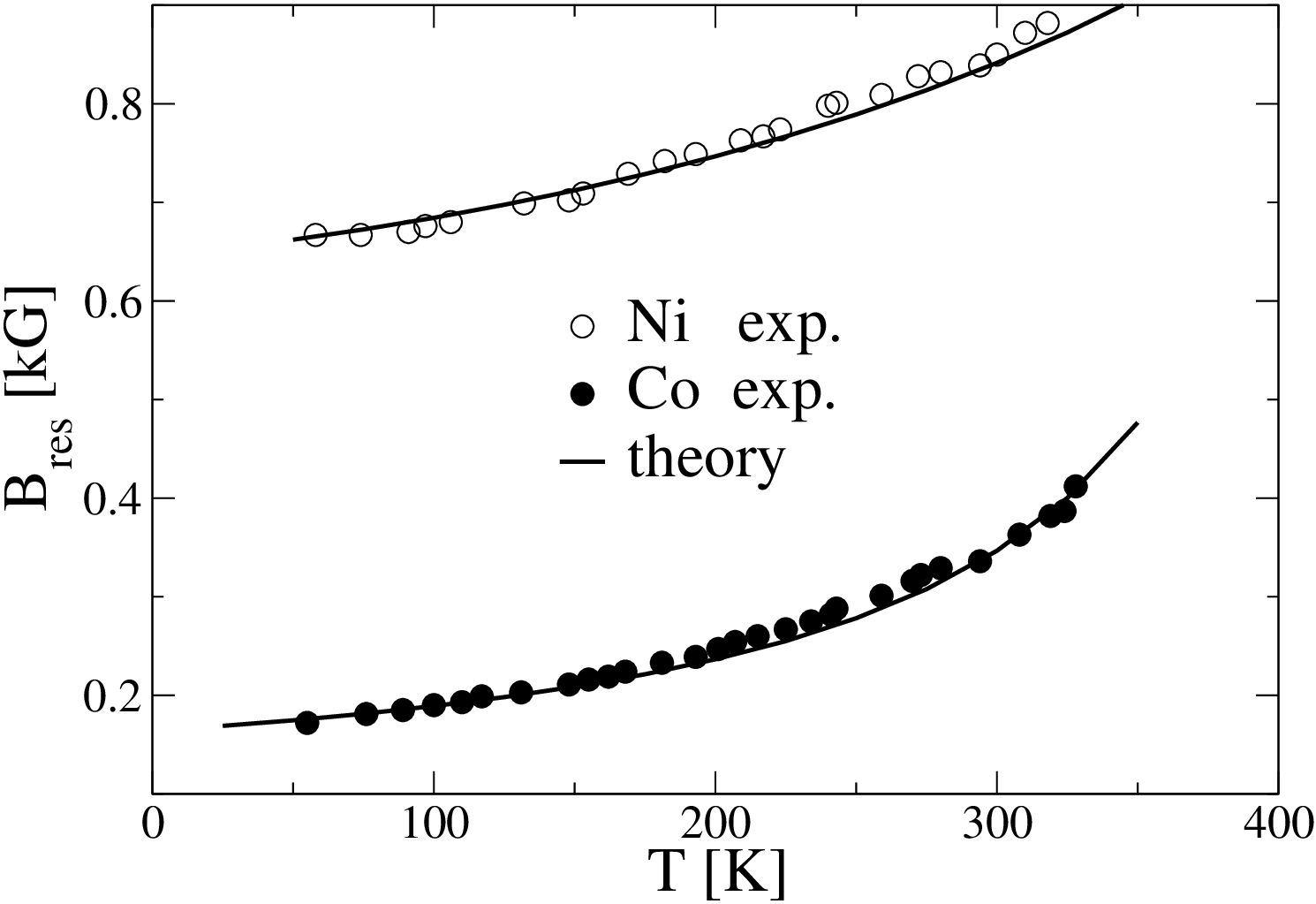}
\caption{Full temperature dependence of the resonance field in the easy direction ($\theta_{B_{0}}=90^{\circ}$). Circles are experimental data, 
lines are from theory with the same parameters as in Fig. 2 and Fig. 3.}
\label{fig4}
\end{figure}

The experimental data have been obtained using FMR measurements \cite{LB03,L03}. This technique probes the uniform
spin wave mode $\omega({\bf q=0})$ of a magnetic sample. An external field is tuned for a given probe frequency 
$\nu_{0}=\omega({\bf q=0})/2\pi$  until resonance occurs at $B_{\rm res}(\theta_{B_{0}})$, with $\theta_{B_{0}}$ being the angle between 
the magnetic field and the normal to the filmplane. This is illustrated in Fig. \ref{fig1}.

Using the temperature dependent effective anisotropy (\ref{Keff}) we can now fit the experimental data. We can thus check if the temperature dependence of the effective 
anisotropy is due to spin wave excitations which are considered explicitly in our model (\ref{hamilt}) or due to other (non-magnonic) effects which would manifest 
themselves in a temperature dependence of the parameters $K_{2}$ and $g_{0}$.

\section{Fit to experiments with Ni and Co films}\label{}

In the following we take $S=1$ for Ni and $S=2.5$ for Co due to different magnetic moments of the two metals \cite{remark}. The Land{$\rm\acute e$} factor is taken as $g_{J}=2.1$ for both Ni and Co. The FMR microwave frequency was set to $9~$GHz.

Fig. \ref{fig2} and \ref{fig3} show the comparison between the $B_{res}(\theta_{B_{0}})$-curves from theory and experimental data for a (subscript denotes the 
number of monolayers) Cu$_9$/Co$_2$/Cu(001) and a Ni$_7$/Cu(001) film system, respectively, at two different temperatures. In both cases the effective anisotropy 
favors the magnetization to lie within the film plane. There is quite good agreement at both temperatures over the whole range of angles $\theta_{B_{0}}$ 
for both films. At a given angle the resonance field increases with temperature.

It is important to note that the choice of the microscopic parameters $K_{2}$ and $g_{0}$ at 
a given temperature cannot be unambiguous as one easily sees from (\ref{Keff}). However we took additionally into account the temperature 
dependence of the resonance field for a fixed angle $\theta_{B_{0}}=90^{\circ}$ as it is shown in Fig. \ref{fig4}. Due to the temperature dependent term which goes 
with $K_{2}$ in (\ref{Keff}), namely $\langle S_{z}^2\rangle(T)$, the ambiguity is removed. Indeed it is still possible to accurately fit the 
experimental results with one set of ($T$-independent) parameters ($K_{2},~g_{0}$) for Ni and Co, respectively, over the whole temperature range. Furthermore 
in both cases the values of $g_{0}S$ lie slightly above the result of an explicit evaluation of this quantity assuming point-like dipoles on the lattice sites 
for the given geometry \cite{L03} ($g_{0}S=3.81\mu_{B}$kG).
The conclusion we can draw is that the temperature dependence of the magnetic anisotropy energy is solely due to spin wave excitations which manifest themselves 
in the $T$-dependence of (\ref{Keff}) rather than due to thermal expansion or phononic interactions. In other words, there is no additional $T$-dependence 
of the parameters $K_{2},~g_{0}$ to be considered in order to describe the non-magnonic effects.

In conclusion we presented a quantum theory for thin metallic films based on a local-moment model with lattice and shape anisotropy. By comparison with FMR 
experiments we found that the temperature dependence of the magnetic anisotropy energy of thin Ni and Co films is exclusively due to magnon excitations rather than 
caused by other structural or phononic effects.

Financial support by the SFB 290 for part of this work is gratefully acknowledged.


\begin{thebibliography}{00}
\bibitem{remark}
These values are larger than the magnetic moments of Ni and Co might suggest. However the crucial quantities in 
our theory are $JS(S+1),~K_{2}(2S-1),~g_{0}S$ and thus scaling $S$ can be absorbed into the other parameters.
\end{thebibliography}
\end{document}